\documentclass[conference]{IEEEtran}
\usepackage{listings}
\usepackage{algorithm}
\usepackage[noend]{algpseudocode}
\usepackage{amsmath}
\usepackage{pdfpages}

\makeatletter
\def\BState{\State\hskip-\ALG@thistlm}
\makeatother

\ifCLASSINFOpdf
\else
\fi
\begin{document}
%
\title{Comparing Dataset Characteristics that Favor the Apriori, Eclat or FP-Growth Frequent Itemset Mining Algorithms}

\author{\IEEEauthorblockN{Jeff Heaton}
\IEEEauthorblockA{College of Engineering and Computing\\
Nova Southeastern University\\
Ft. Lauderdale, FL 33314\\
Email: jeffheaton@acm.org}}


%


\maketitle

\begin{abstract}
Frequent itemset mining is a popular data mining technique.  Apriori, Eclat, and FP-Growth are among the most common algorithms for frequent itemset mining.  Considerable research has been performed to compare the relative performance between these three algorithms, by evaluating the scalability of each algorithm as the dataset size increases.  While scalability as data size increases is important, previous papers have not examined the performance impact of similarly sized datasets that contain different itemset characteristics. 

This paper explores the effects that two dataset characteristics can have on the performance of these three frequent itemset algorithms.  To perform this empirical analysis, a dataset generator is created to measure the effects of frequent item density and the maximum transaction size on performance.  The generated datasets contain the same number of rows.  This provides some insight into dataset characteristics that are conducive to each algorithm. The results of this paper's research demonstrate Eclat and FP-Growth both handle increases in maximum transaction size and frequent itemset density considerably better than the Apriori algorithm. 
\end{abstract}


%
\IEEEpeerreviewmaketitle

\section{Introduction}

The research covered by this paper determines how the characteristics of a dataset might affect the performance of the Apriori, Eclat, and FP-Growth frequent itemset mining algorithms.  These algorithms have several popular implementations\cite{han2007,hall2009weka,Hahsler2005}.   The goal of this research is to determine the effects of basket size and frequent itemset density on the Apriori, Eclat, and FP-Growth algorithms. The research determined that these two dataset characteristics have a significant impact on performance of the algorithms.

Most research into frequent itemset mining focuses upon the performance differences between frequent itemset algorithms on a single dataset\cite{burdick2001mafia}. The effects of hyper-paramaters, such as minimum support, upon the performance of frequent itemset mining algorithms has also been explored\cite{zheng2001real}.  Some papers make use of common datasets from the UCI Machine Learning Repository\cite{AsuncionNewman2007}. Many papers make use of the IBM Quest Synthetic Data Generator\cite{pit11} or some variant of it.  Our paper makes use of a Python-based generator that is based on IBM's work\cite{jeffpapers}.

This research evaluates the performance of the Apriori, Eclat and FP-Growth frequent itemset mining algorithms implemented by Christian Borgelt in 2012\cite{borgelt2012frequent}.  Though, association rule mining is a similar algorithm, this research is limited to frequent itemset mining.  By limiting the experimentation to a single implementation of frequent itemset mining this research is able to evaluate how the characteristics of the dataset affect the performance of these algorithms.

\section{Frequent Itemset Mining}

Frequent itemset mining was introduced as a means to find frequent groupings of items in a database containing baskets/transactions of these items\cite{agrawal1993mining}. The database is composed of a series of baskets that are analogous to orders placed by customers.  These orders are individual baskets that are made up of some number of items.  Companies, such as Amazon, Netflix and other online retailers, make use of frequent itemsets to suggest additional items that a consumer might want to purchase, based on their past purchasing history and the history of others with similar baskets\cite{leskovec2014mining}.
   The following data show baskets that might be used for frequent itemset mining, where each line represents a single basket of items.  

\begin{lstlisting}
[mp3player usb-charger book-dct book-ths]
[mp3player usb-charger]
[usb-charger mp3player book-dct book-ths]
[usb-charger]
[book-dct book-ths]
\end{lstlisting}

From the above baskets several frequent itemsets can be defined.  These are sets of items that frequently occur together, some of which are:

\begin{lstlisting}
[mp3player usb-charger]
[book-dct book-ths]
...
\end{lstlisting}

A simple visual analysis of the data show that the items mp3-player and usb-charger frequently occur together.  Likewise, book-dct and book-ths also frequently occur together.  Frequent itemset algorithms make use of a variety of statistics to determine which itemsets to include.

\section{Survey of Apriori, Eclat and FP-Growth}

There are a variety of different algorithms that are used to mine frequent itemsets.  First, a simple naive brute-force algorithm to build frequent itemsets will be evaluated.  This paper shows how Apriori, Eclat and FP-Growth address some of the shortcomings of the naive algorithm. 

All four algorithms must calculate statistics about itemsets that might ultimately be included in the final collection of frequent itemsets.  One statistic that is common to all four of these algorithms is support.  The support of a candidate frequent itemset is the total count of how many of the database baskets support that candidate.  A basket is said to cover a candidate itemset if the candidate is a subset or equal to the basket.  Support is sometimes expressed as a percent of the total number of baskets in the database ($N$) that cover a candidate itemset ($X$). The following formula calculates the support percentage of a candidate itemset:

\begin{equation}
\text{supp}(X) = \frac{X count}{N}
\end{equation}

This equation can be applied to calculate the support for $\{$mp3-player usb-charger$\}$ from the previously presented set of baskets.

\begin{equation}
\text{supp}( \{\text{mp3-player} \  \text{usb-charger}\})=  \frac{3}{5}=0.6
\end{equation}

 The support statistic of 0.6 indicates that 60\% of the five baskets contain the candidate itemset $\{$mp3-player usb-charger$\}$.  Most frequent itemset algorithms accept a minimum support parameter to filter out less common itemsets.
 
\section{Naive Algorithm for Finding Frequent Itemsets}
 
It is not difficult to extract frequent itemsets from basket data.  It is, however, difficult to do so efficiently.  For the algorithms presented here, let $\mathcal{J}$ represent a set of items, likewise, let $\mathcal{D}$ represent a database of baskets that are made up of those items.  Algorithm 1 is a summarization of the naive frequent itemset algorithm provided by Garcia-Molina, Ullman, and Widom\cite{garcia2008database}.
 
\begin{algorithm}
\caption{Naive Frequent Itemset Algorithm}\label{naive}
\begin{algorithmic}[1]
\BState {\textbf{INPUT:}} A file $\mathcal{D}$ consisting of baskets of items.
\BState {\textbf{OUTPUT:}} The sets of itemsets $F_1$, $F_2$, \ldots, $F_q$, where $F_i$ is the set of all itemsets of size \textit{I} that appear in at least \textit{s} baskets of $\mathcal{D}$.
\BState {\textbf{METHOD:}}
\State $R \gets \text{integer array, all item combinations in $\mathcal{D}$,  of size $2^{|\mathcal{D}|}$}$
\For{$n \gets 1 \ TO \ |\mathcal{D}|$}
\State $F \gets$ \text{all possible set combinations from $D_n$}
\State{Increase each value in R, corresponding to each in $F[]$}
\EndFor
\Return{all itemsets, with $R[]\ge s$} 
\end{algorithmic}
\end{algorithm}
 
The naive algorithm simply generates all possible itemsets, counts their support, and then discards all itemsets below some threshold level of support.  The constant $S$ or $\sigma$ typically represents the support threshold.  Computing all possible itemsets is only an $O(N)$ magnitude operation in all cases, where $N$ is the number of baskets in the database.  However, the naive algorithm would also need $2^i$ memory cells to store all of these itemsets as the counts are generated, where $i$ is the number of individual items.  These memory cells would typically be 32 or 64-bit integers.  This memory requirement means that the naive algorithm is impossible for anything but a trivial number of individual items.  A computer with 128GB of available RAM would theoretically only be able to calculate 34 items, when using a 64-bit integer to hold the counts. When it is considered that $N$ might be the total count of distinct items for sale by a retailer such as Walmart or Amazon it is obvious that the naive approach is not useful in practice. 

\section{Naive Algorithm Example}
This section demonstrates how the naive algorithm would handle the example basket set given earlier in this paper.  The total number of items contained in database, $|\mathcal{J}|$ is equal to four.  Four items can be arranged a total of $|\mathcal{J}|^2$, or 16 different ways.  However, because one of these frequent itemsets is the empty set, only the following 15 candidate itemsets are considered:

\begin{lstlisting}
[book-ths] 
[book-dct] 
[book-dct, book-ths] 
[usb-charger]
[usb-charger, book-ths] 
[usb-charger, book-dct] 
[usb-charger, book-dct, book-ths] 
[mp3player]
[mp3player, book-ths] 
[mp3player, book-dct] 
[mp3player, book-dct, book-ths] 
[mp3player, usb-charger]
[mp3player, usb-charger, book-ths] 
[mp3player, usb-charger, book-dct] 
[mp3player, usb-charger, book-dct, book-ths]
\end{lstlisting}

The above itemsets are considered candidate frequent itemsets because it has not yet been determined if all of these candidates will be included in the final list of frequent itemsets.  
Once the candidate itemsets have been determined, the naive algorithm will pass over all baskets and count the support for each of the candidate itemsets.  Candidate itemsets that are below the required support $S$ will be purged.  The naive algorithm would calculate support for each candidate as follows:
\begin{lstlisting}
[book-ths]; s = 3
[book-dct]; s = 3
[book-dct book-ths]; s = 3
[usb-charger]; s = 4
[usb-charger, book-ths]; s = 2
[usb-charger, book-dct]; s = 2 
[usb-charger, book-dct, book-ths]; s = 2 
[mp3player]; s = 3
[mp3player, book-ths]; s = 2
[mp3player, book-dct]; s = 2
[mp3player, book-dct, book-ths]; s = 2
[mp3player, usb-charger]; s = 3
[mp3player, usb-charger, book-ths]; s = 2
[mp3player, usb-charger, book-dct]; s = 2
[mp3player, usb-charger, book-dct, 
  book-ths]; s = 2
\end{lstlisting}
It is necessary to store a count for every possible itemset when using the naive algorithm.  Of course, once the support counts are determined, many of the frequent itemsets will be purged.  Nevertheless, the fact that these values must be kept while the database is scanned for support means the naive algorithm requires considerable memory.

\section{Apriori Algorithm for Finding Frequent Itemsets}
Agrawal and Srikant initially introduced the Apriori algorithm to provide performance improvements over a naive itemset search\cite{agrawal1994fast}. Apriori algorithm has been around almost as long as the concept of frequent itemsets and is very popular.  The naive algorithm is a theoretical concept and is not used in practice.  Aprioiri has become the classic implementation of frequent itemset mining.  Aprioiri, as defined by Goethals (2003) is presented as Algorithm 2\cite{goethals2003survey}.

\begin{algorithm}
\caption{Apriori Frequent Itemset Algorithm}\label{apriori}
\begin{algorithmic}[1]
\BState {\textbf{INPUT:}} A file $\mathcal{D}$ consisting of baskets of items, a support threshold $\sigma$.
\BState {\textbf{OUTPUT:}} A list of itemsets $\mathcal{F}(\mathcal{D},\sigma)$.
\BState {\textbf{METHOD:}}

\State {$C_1 \gets \{ \{i\} | i \in \mathcal{J} \} $}
\State $k \gets 1$
\While{ $C_k \neq \{\}$ }
\\ \# {Compute the supports of all candidate itemsets}
  \For { all transactions $\{tid,I\} \in \mathcal{D} $ }
    \For {all candidate itemsets $X \in C_k$} 
      \If{ $X \subseteq I$ }
        \State{ $X.support$++}
      \EndIf
    \EndFor
  \EndFor
\\ \#  {Extract all frequent itemsets}
  \State{ $F_k = \{ X | X.support > \sigma \} $}
\\ \# {Generate new candidate itemsets}
  \For { all $X,Y \in Fi, X[i] = Y[i]$ for $1 \leq i \leq k-1, 
                         \text{and} \ X[k] < Y[k] $ }
  \State{ $I = X \cup \{Y[k]\}$}
    \If {$\forall J \subset I, |J| = k : J \in F_k $}
      \State{$ C_{k+1} \gets C_{k+1} \cup I $}
    \EndIf
  \EndFor
  \State{ $k$++}
\EndWhile

\end{algorithmic}
\end{algorithm}

Apriori is based on the hierarchical monotonicity of frequent itemsets between their supersets and subsets. As implied by monotonicity, a subset of a frequent itemset must also be frequent.  Likewise, a superset of an infrequent itemset must also be infrequent\cite{agrawal1994fast}.   This allows the Apriori algorithm to be implemented as a breadth-first search.  Papers by Goethals (2003) and others do not represent Apriori's performance in terms of big-O notation\cite{goethals2003survey}.  This is likely due to the fact that Apriori's outer loops are bounded by the number of common prefixes and not some easily determined constants such as the number of items or the length of the dataset. Papers describing Apriori, Eclat, and FP-Growth rely on empirical comparison of algorithms rather than big-O analysis. However, analysis covered later in this paper does allow these three algorithms to be expressed in big-O based on average basket size and frequent itemset density.

Aprioiri first builds a list of all singleton itemsets with sufficient support.  Building on the monotonicity principle, the next set of candidate frequent itemsets is built of combinations of the singleton itemsets.  This process continues until the maximum length specified for frequent itemsets is reached.  The evaluations performed by this research did not impose this maximum length.

The primary issue with the Apriori algorithm is that it is necessary to perform a scan of the database at every level of the breadth-first search.  Additionally, candidate generation can lead to a great number of subsets and can become a significant memory requirement.  Deficiencies in the Apriori algorithm led to the development of other, more efficient, algorithms, such as Eclat and FP-Growth.

\section{Apriori Algorithm Example}

This section will demonstrate how the Apriori algorithm handles the basket set given earlier in this paper.  The Apriori algorithm performs a breadth first search of the itemsets.  Figure 1 shows a segment of this search, for the items usb-cable, mp3-player, and book-dct. 

\begin{figure}[h]
   \centering      
   \includegraphics[width=8cm]{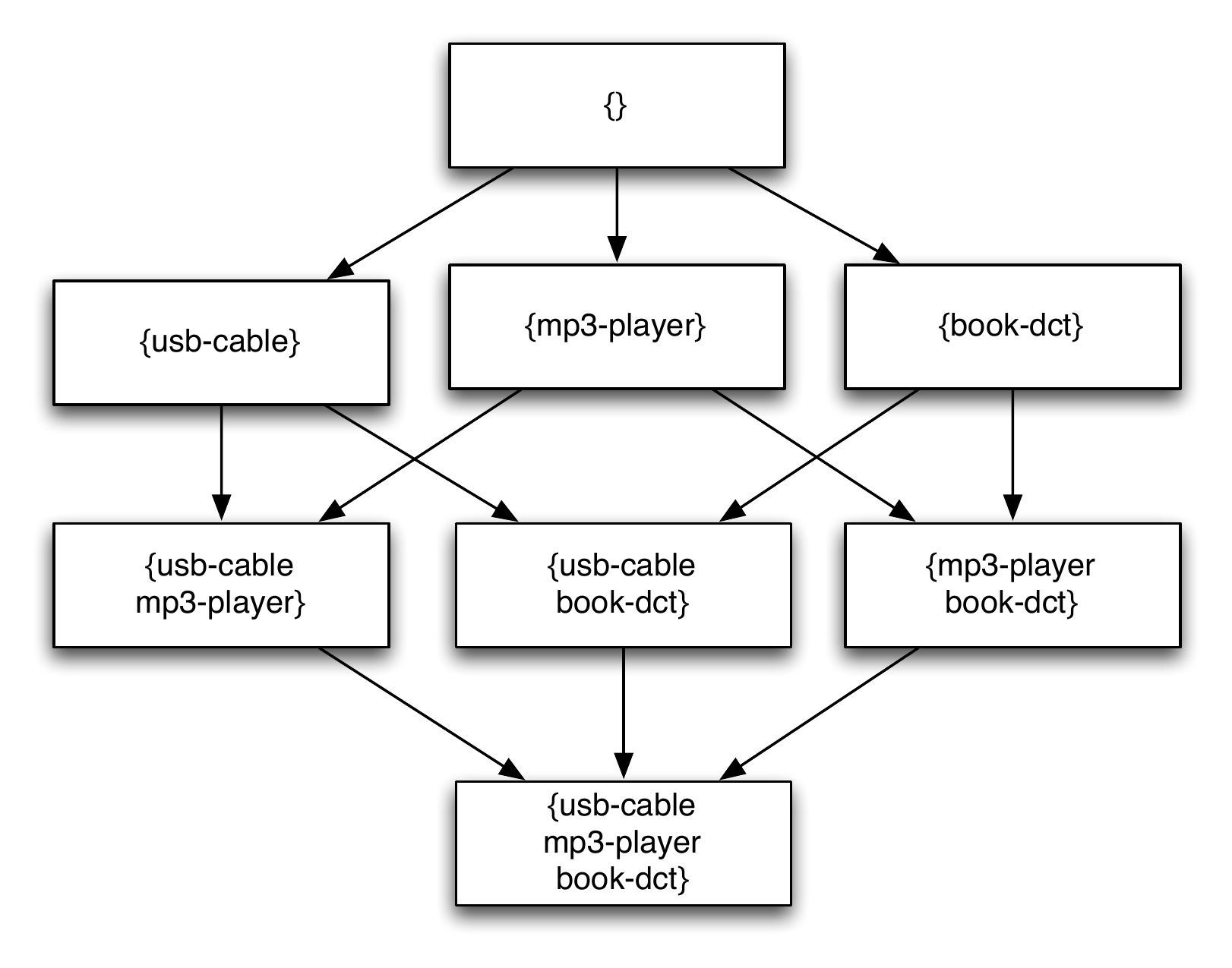}      
 \caption{Apriori Breadth-First Search}
 \label{fig:apriori}
\end{figure}

The candidate set starts empty, and begins by adding all singleton itemsets that have sufficient individual support.  For simplicity, it is assumed that only usb-cable, mp3-player, and book-dct have sufficient support.  The next layer is built of combinations of the previous layer that had sufficient support.  For simplicity, it is also assumed that all three combinations had sufficient support.  Finally, a triplet itemset with all three items is evaluated.

\section{Eclat Algorithm for Finding Frequent Itemsets}

Eclat was introduced by Zaki, Parthasarathy, Ogihara, and Li in 1997\cite{zaki1997new}.  Eclat is an acronym for Equivalence Class Clustering and bottom up Lattice Traversal. The primary difference between Eclat and Apriori is that Eclat abandons Apriori's breadth-first search for a recursive depth-first search.  Eclat, as defined by Goethals (2003) is presented as Algorithm 3\cite{goethals2003survey}.

\begin{algorithm}
\caption{Eclat Frequent Itemset Algorithm}\label{eclate}
\begin{algorithmic}[1]
\BState {\textbf{INPUT:}} A file $\mathcal{D}$ consisting of baskets of items, a support threshold $\sigma$, and an item prefix $I$, such that $I \subseteq \mathcal{J}$. 
\BState {\textbf{OUTPUT:}} A list of itemsets $\mathcal{F}[I](\mathcal{D},\sigma)$ for the specified prefix.
\BState {\textbf{METHOD:}}

\State { $\mathcal{F}[i] \gets \{\}$}
\For{ all $i \in \mathcal{J}$ occurring in $\mathcal{D}$ }
  \State{ $ \mathcal{F}[I] := \mathcal{F}[I] \cup \{ I \cup \{i\} \} $ }
  \\ \# { Create $\mathcal{D}_i$}
  \State{$\mathcal{D}_i \gets \{\} $} 
  \For {all $j \in \mathcal{J}$ occurring in $\mathcal{D}$ such that $j > i$ }
    \State{ $C \gets$ cover($\{i\}$) $\cap$ cover($\{j\})$ }
    \If{$ |C| \geq \sigma$} 
      \State{ $\mathcal{D}_i \gets \mathcal{D}_i \cup \{j,C\}$}
    \EndIf
  \EndFor
  \\ \# {Depth-first recursion}
  \State{ Compute $\mathcal{F}[I \cup {i}](\mathcal{D}_i,\sigma) $}
  \State{ $\mathcal{F}[I] := \mathcal{F}[I] \cup \mathcal{F}[I \cup {i}]$ }
\EndFor

\end{algorithmic}
\end{algorithm}

The input parameters to Eclat are slightly different than Apriori in that a prefix $I$ is provided.  This prefix specifies the prefix pattern that must be present in any itemsets found by the call to Eclat.  This change allows a depth-wise recursive building of the itemsets.  The initial call to Eclat uses an $I$ value of $\{\}$, meaning that no specific prefix is required.  This initial call would find all single-item frequent itemsets.  The Apriori algorithm would then recursively call itself, each time increasing $I$ by adding itemsets that contain the value $I$ that the function was called with, but are one item longer. This process would continue until the value of $I$ had grown to sufficient length that the algorithm has traversed to baskets of all lengths.  Like Apriori, Eclat is not usually expressed in big-O terms; however, results obtained from this research's experimentation show how frequent itemset density and basket allow these algorithms to be expressed in terms of big-O computational cost.

There are several different methods for storing the support values in the recursive Eclat algorithm.  The most common approach is to use a structure called a trie.  This is the approach used by Borgelt (2012) to implement the versions of Apriori, Eclat and FP-Growth investigated in this research paper\cite{borgelt2012frequent}.  A trie graph always contains an empty root node.  As itemsets are encountered, they are added to the trie by inserting a node for each item that makes up the itemset.  The left-most item corresponds to a child of the root node.  The second item corresponds to a child of the first item of this frequent set.   No parent would ever have more than one child of the same item name; however, an item name may appear at multiple locations in the trie.  

The trie is generated so that the algorithm can quickly find the support of an itemset by traversing the trie as the items in the set are read left-to-right.  The node that contains the right-most item contains the support for that itemset.  As the algorithm processes the database the trie is traversed looking for each itemset discovered.  Nodes are created, if necessary, to fill out the trie to hold all itemsets.  If the nodes already exist, the node for the right-most item in the itemset has its support increased. New nodes start with a support of 1.  This allows Eclat to use less memory than Apriori, because the core branches of the trie allow heavily used subsets to be stored only once.  Theoretically, a trie could be used with Apriori, however, the breadth-first nature of Apriori would typically require too much memory.

\section{Eclat Algorithm Example}

This section will demonstrate how the Eclat algorithm would handle the basket set given earlier in this paper.  A portion of the trie built by Eclat is shown in Figure 2.

\begin{figure}[h]
   \centering      
   \includegraphics[width=8cm]{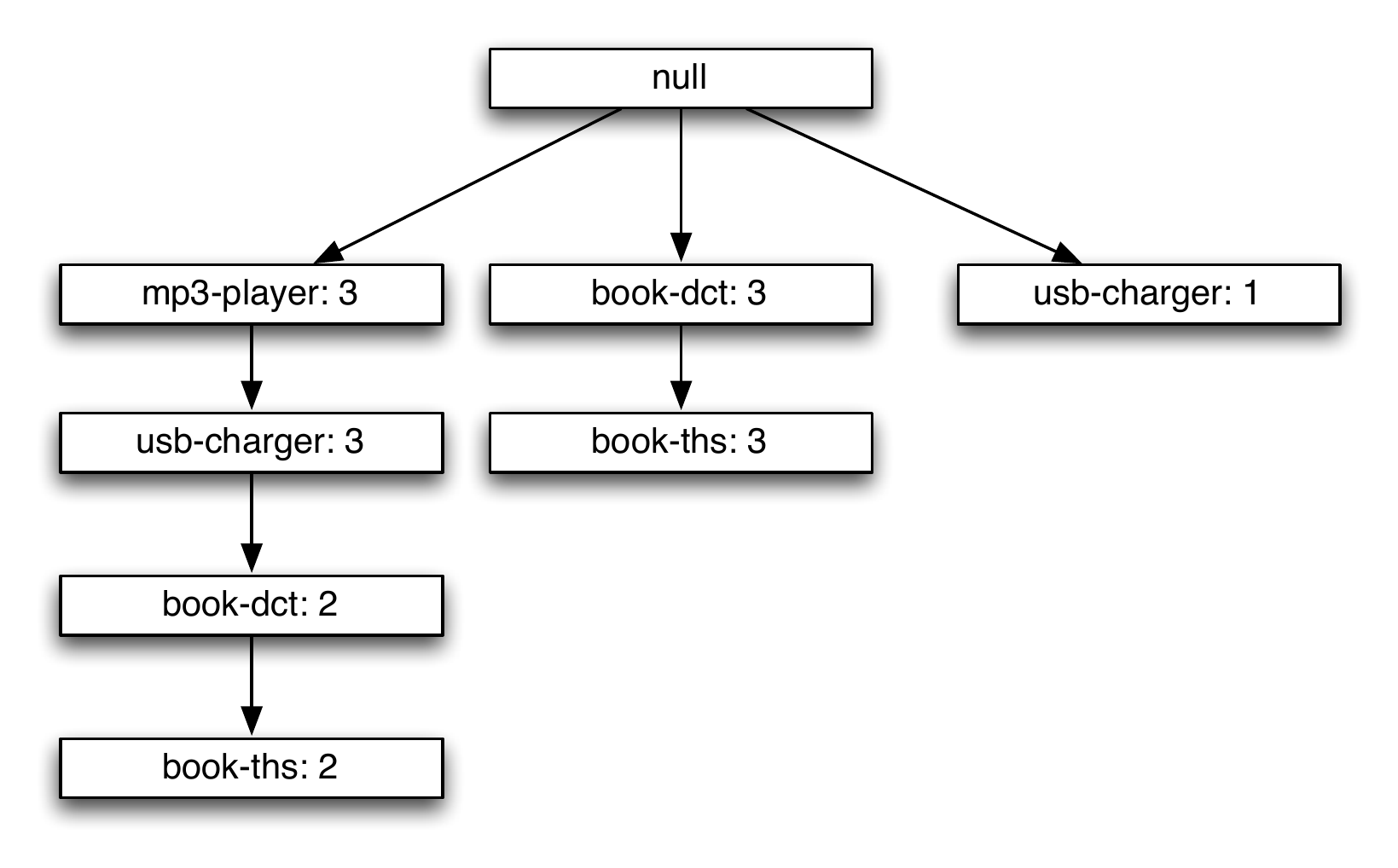}      
 \caption{A Trie Used by Eclat }
 \label{eclat}
\end{figure}

The above trie portion encodes a total of 7 different frequent itemset's support values.  To find a particular frequent itemset's support simply traverse the graph, following the items from left to right.  The frequent itemset mp3-player, usb-charger would have a support value of 3.  Similarly, the frequent itemset mp3-player, usb-charger, book-dct would have a support value of 2.  Once the algorithm completes, the trie is traversed, and all frequent itemsets are extracted from it.

\section{FP-Growth Algorithm for Finding Frequent Itemsets}

Frequent pattern growth (FP-Growth) was introduced by Han, Pei, and Yin in 2000 to forego candidate generation altogether\cite{han2000mining}.  This is done by using a trie to store the actual baskets, rather than storing candidates like Apriori and Eclat do.  Aprori is very much a horizontal, breadth-first, algorithm.  Similarly, Eclat is very much a vertical, depth-first, algorithm. The trie structure of FP-Growth provides a vertical view of the data.  However, FP-Growth also adds a header table for every individual item that has support above the threshold support level.  This header table contains a linked-list through the trie to connect every node of the same type.  The header table gives FP-Growth a horizontal view of the data, in addition to the vertical view provided by the trie.

The algorithm for FP-Growth is similar to Eclat in that it was not expressed in terms of big-O analysis.  FP-Growth, as defined by Goethals, is presented as Algorithm 4\cite{goethals2003survey}.

\begin{algorithm}
\caption{FP-Growth Frequent Itemset Algorithm}\label{FP-Growth}
\begin{algorithmic}[1]
\BState {\textbf{INPUT:}} A file $\mathcal{D}$ consisting of baskets of items, a support threshold $\sigma$, and an item prefix $I$, such that $I \subseteq \mathcal{J}$. 
\BState {\textbf{OUTPUT:}} A list of itemsets $\mathcal{F}[I](\mathcal{D},\sigma)$ for the specified prefix.

\State {$\mathcal{F}[i] \gets \{\}$ }
\For {all $i \in \mathcal{J}$ occurring in $\mathcal{D}$ }
  \State {$\mathcal{F}[I] \gets \mathcal{F}[I] \cup \{ I \cup \{i\}$ }
\\ \# Create $\mathcal{D}_i$
  \State {$\mathcal{D}_i \gets \{\} $}
  \State {$H \gets \{\}$}
  \For {all $j \in \mathcal{J}$ occurring in $\mathcal{D}$ such that $j > i$ }
    \If {support($I \cup \{i,j\}$) $\geq \sigma $}
      \State{ $H \gets H \cup  \{j\}$ }
    \EndIf
  \EndFor

  \For {all $(tid,X) \in \mathcal{D}$  with $I \in X$ }
    \State {$\mathcal{D}_i \gets \mathcal{D}_i \cup \{(tid,X \cap H)\}$}
  \EndFor
  
  \\ \# Depth-first recursion
  \State {Compute $\mathcal{F}[I \cup \{i\}](\mathcal{D}_i,\sigma) $}
  \State { $\mathcal{F}[I] \gets \mathcal{F}[I] \cup \mathcal{F}[I \cup \{i\}]$}
\EndFor

\end{algorithmic}
\end{algorithm}

\section{FP-Growth Algorithm Example}

This section will demonstrate how the FP-Growth algorithm would handle the basket set given earlier in this paper.  Figure 3 shows a portion of the FP-Growth trie and the header table generated for the earlier example data.

\begin{figure}[h]
   \centering      
   \includegraphics[width=9cm]{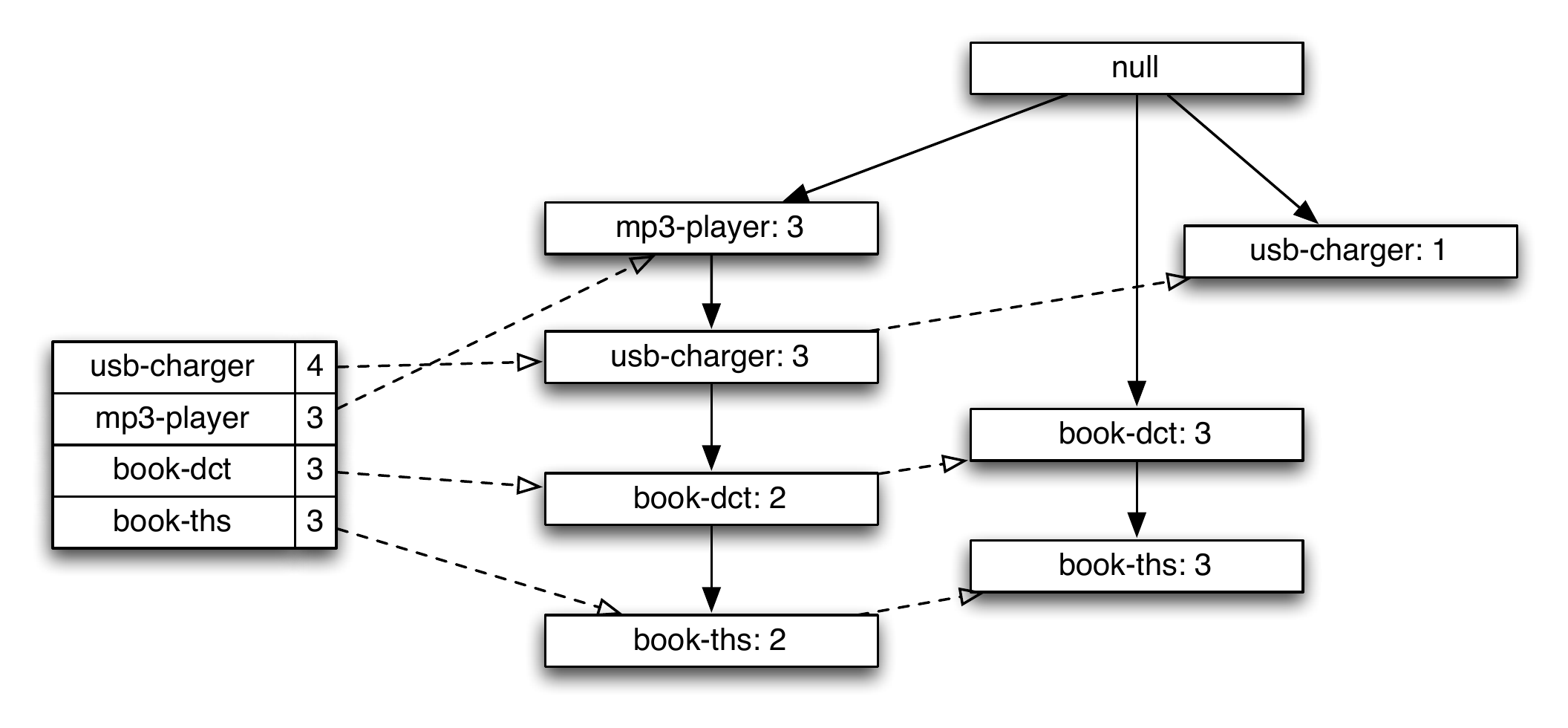}      
 \caption{FP-Growth Trie and Header Table }
 \label{fig:fp-growth}
\end{figure}

This figure demonstrates the horizontal and vertical nature of the FP-Growth algorithm.  The trie, on the right, holds the encoded baskets, along with their supports.  The header, on the left, holds the items and provides horizontal access to the data.

\section{Empirical Comparison of Apriori, Eclat and FP-Growth}

There are a number of papers that compare the computational performance of Apriori, Eclat and FP-Growth.  These papers are typically focused primarily on comparing the differences between the algorithms on one or more datasets and at different support thresholds.  Papers by Borgelt (2012) and Goethals (2003) are examples of papers that compare various implementations of Apriori, Eclat and FP-Growth\cite{borgelt2012frequent,goethals2003survey}.

This paper attempts a different approach.  The goal of this paper is to see what effect the dataset has on the algorithm.  The average basket size and frequent item density will be used as independent variables to evaluate total processing time as the dependent variable.  Apriori, Eclat and FP-Growth will each be evaluated independently to see which performs best at different basket sizes and frequent itemset densities.  Performance will be measured as a shorter total runtime.  This paper focuses on a single implementation of these algorithms using the implementations of Apriori, Eclat and FP-Growth by Borgelt in 2012\cite{borgelt2012frequent}.

\section{Dataset Generation}

Generated datasets are used to perform this evaluation.  This generated data allows the two independent variables to be adjusted to create a total of 20 different datasets to perform the evaluations.  The datasets were generated using a simple Python script created for this paper that can be found on GitHub\cite{jeffpapers}.  This Python script accepts the following parameters to specify the dataset to produce:

\begin{itemize}
\item	Transaction/Basket count: 10 million default
\item	Number of items: 50,000 default
\item	Number of frequent sets: 100 default
\item	Max transaction/basket size: independent variable, 5-100 range
\item	Frequent set density: independent variable, 0.1 to 0.8 range
\end{itemize}

While basket count, number of frequent sets, and number of items can easily be varied in the script, for the purposes of this paper they will remain fixed at the above values.  Through informal experimentation it was determined that the basket count only had a small positive correlation to processing time.  The number of items did not appear to have a meaningful impact on processing time when varied in isolation.  It was observed that the strongest correlation to processing time was through variation of the maximum basket size and frequent set density.

The following listing shows the type of data generated for this research.  Here an example file was created with 10 baskets, out of 100 items, 2 frequent itemsets, maximum basket size of 10, and a density of 0.5.

\begin{lstlisting}
I36 I94
I71 I13 I91 I89 I34
F6 F5 F3 F4
I86
I39 I16 I49 I62 I31 I54 I91
I22 I31
I70 I85 I78 I63
F4 F3 F1 F6 F0 I69 I44
I82 I50 I9 I31 I57 I20
F4 F3 F1 F6 F0 I87
\end{lstlisting}

As you can see from the above file, the items are either prefixed with ``I'' or ``F''.  The ``F'' prefix indicates that this line contains one of the frequent itemsets.  Items with the ``I'' prefix are not part of an intentional frequent itemset.  Of course, ``I'' prefixed items might form frequent itemsets, as they are uniformly sampled from the number of items to fill out non-frequent itemsets.  Each basket will have a random size chosen, up to the maximum basket size.  The frequent itsemset density specifies the probability of each line containing one of the intentional frequent itemsets.  Because a density of 0.5 was used, approximately half of the lines above contain one of the two intentional frequent itemsets.  A frequent itemset line may have additional random ``I'' prefixed items added to cause the line to reach the randomly chosen length for that line.  If the chosen frequent itemset does cause the generated line to exceed its randomly chosen length, no truncation will occur.  The intentional frequent itemsets are all chosen to be less than or equal to the maximum basket size. 

\section{Effects of Dataset Density}

Dataset density specifies the percentage of baskets that intentionally contain frequent itemsets.    As frequent itemset density increases so does the processing time of Apriori, Eclat and FP-Growth as shown by Figure 4.

\begin{figure}[h]
   \centering      
   \includegraphics[width=9cm]{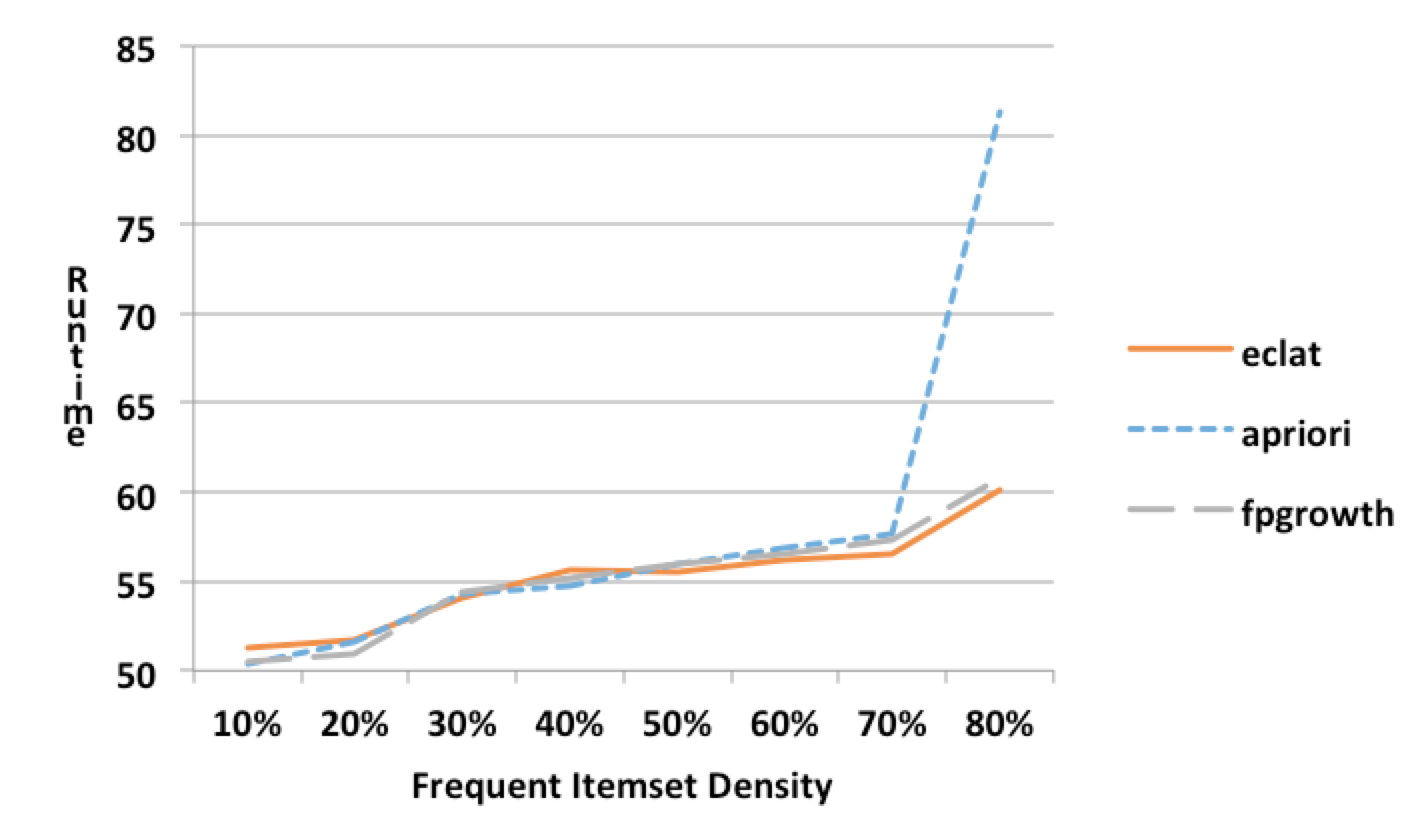}      
 \caption{Frequent Itemset Density's Effect on Runtime (seconds)}
 \label{fig:density}
\end{figure}
 
This chart shows the results of running 10 million baskets, with an average basket size of 50, at various densities.  The Eclat and FP-Growth algorithms both show very similar growth as frequent itemset density increases.  The Apriori algorithm also performs very similarly to Eclat and FP-Growth, until the density surpasses 70\%.  As previously mentioned in this paper, Apriori has considerably larger memory needs than the other algorithms.  At 70\% Apriori had allocated all of the test machine's 8 gigabytes of RAM. This made swapping to physical storage necessary and had a drastic impact on the algorithm's runtime.  It is also interesting to note that Eclat is marginally ahead of FP-Growth at low densities.  This ranking reverses at higher densities.  Between 10\% and 70\% all three algorithms exhibit approximately an $O(N log N)$ complexity.  Beyond 70\% Apriori approached $O(N^2)$ and worse complexity, where $N$ is the number of actual frequent items in the database.

\section{Effects of Basket Size}

Basket size specifies the maximum number of items per basket line. Larger basket sizes mean that the frequent itemsets will also be larger.  This increases the sizes of the data structures used to hold these itemsets.  These larger data structures require more memory for storage and greater processing time to traverse them.  Figure 5 illustrates the effect of increasing transaction sizes on the performance of the three algorithms.

\begin{figure}[h]
   \centering      
   \includegraphics[width=9cm]{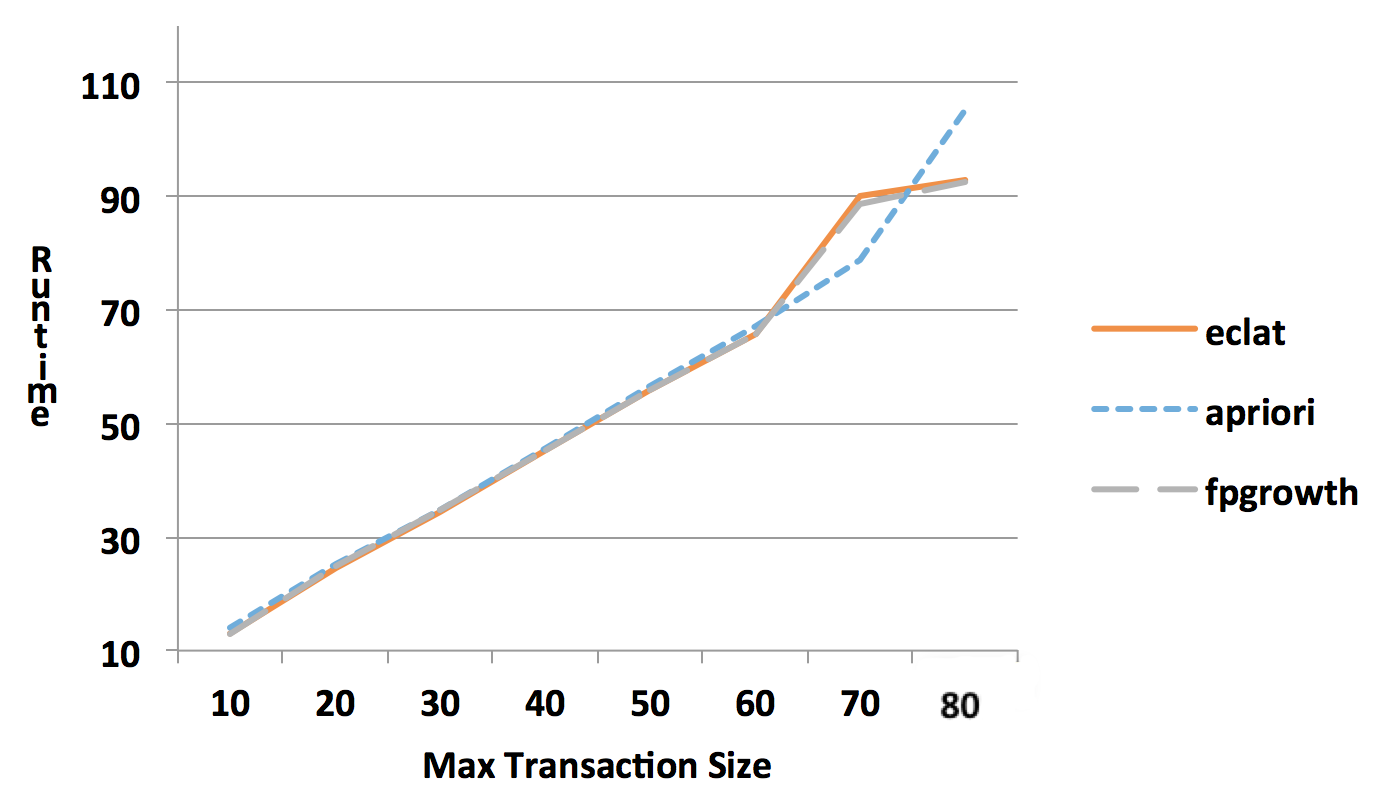}      
 \caption{Maximum Basket Size's Effect on Runtime (seconds)}
 \label{fig:basket}
\end{figure}

This chart shows the results of running 10 million baskets, with a frequent itemset density of 50\%, at various max basket sizes.  The three algorithms show almost exactly the same $O(N)$ performance for basket sizes up through 60.  Once above 60, Apriori seems to grow much quicker than the other two.  This is possibly because of the increased memory used by Apriori.  Interestingly, Apriori performed the best between 60-70 maximum transaction sizes.  Further research is needed to determine why Apriori is superior in this small range.

\section{Conclusions}

Apriori is an easily understandable frequent itemset mining algorithm.  Because of this, Apriori is a popular starting point for frequent itemset study.  However, Apriori has serious scalability issues and exhausts available memory much faster than Eclat and FP-Growth.  Because of this Apriori should not be used for large datasets.  

Most frequent itemset applications should consider using either FP-Growth or Eclat.  These two algorithms performed similarly for this paper's research, though FP-Growth did show slightly better performance than Eclat.  Other papers also recommend FP-Growth for most cases\cite{borgelt2012frequent}.
Frequent itemset mining is an area of active research. New algorithms, as well as modifications of existing algorithms are often introduced.  For an application where performance is critical, it is important to evaluate the dataset with newer algorithms as they are introduced, and shown to have better performance than FP-Growth or Eclat.

\bibliographystyle{IEEEtran}
\bibliography{jheaton_freqitemset}

\end{document}